\documentclass[12pt,preprint]{aastex}
\newcommand \kms{km~s$^{-1}$} 
  
\newcommand{\hi}{H\,{\sc i}}

\begin{document}

\title{A Neutral Hydrogen Survey of the Large Magellanic Cloud:\\
Aperture Synthesis and Multibeam Data Combined}

\author{Sungeun Kim\altaffilmark{1}, Lister Staveley-Smith\altaffilmark{2}, 
Michael A. Dopita\altaffilmark{3}, Robert J. Sault\altaffilmark{2},\\
Kenneth C. Freeman\altaffilmark{3}, Youngung Lee\altaffilmark{4}, and You-Hua Chu\altaffilmark{5}}

\altaffiltext{1}{University of Massachusetts, Astronomy Department, 710 N. Pleasant St., Amherst, MA 01003;
 email: skim@daisy.astro.umass.edu}
\altaffiltext{2}{Australia Telescope National Facility, 76 Epping, NSW 2121, 
Australia; email:lstavele@atnf.csiro.au,rsault@atnf.csiro.au}
\altaffiltext{3}{Mount Stromlo and Siding Spring Observatories, PO Box, 
Canberra, ACT2611, Australia; email:mad@mso.anu.edu.au; kcf@mso.anu.edu.au}
\altaffiltext{4}{Korea Astronomy Observatory, Taeduk Radio Astronomy Observatory, 
Whaam-dong 61-1, Yusong-gu, Taejon,  305-348, Korea; email:yulee@trao.re.kr}
\altaffiltext{5}{Astronomy Department, University of Illinois at 
Urbana-Champaign, 1002 West Green st., IL61801; email:chu@astro.uiuc.edu}

\begin{abstract}

  Recent \hi\ surveys of the Large Magellanic Cloud (LMC) with the
  Australia Telescope Compact Array (Kim et al. 1998) and the Parkes
  multibeam receiver (Staveley-Smith et al. 2003) have focussed,
  respectively, on the small-scale ($<20\arcmin$) structure of the
  interstellar medium (ISM) and the large-scale ($>1\arcdeg$)
  structure of the galaxy. Using a Fourier-plane technique, we have
  merged both data sets providing an accurate set of images of the LMC
  sensitive to structure on scales of 15 pc (for an LMC distance of 50
  kpc) upwards. The spatial dynamic range (2.8 orders of magnitude),
  velocity resolution (1.649 \kms), brightness temperature sensitivity
  (2.4 K) and column density sensitivity ($8.9\times 10^{18}$ cm$^{-2}$ 
  per 1.649 km s$^{-1}$ channel) allow for the studies of phenomena
  ranging from the galaxy-wide interaction of the LMC with its close
  neighbours to the small-scale injection of energy from supernovae
  and stellar associations into the ISM of the LMC. This paper
  presents the merged data and the size spectrum of \hi\ clouds which 
  is similar to the typical size spectrum of the holes and shells in the
  \hi\ distribution.

\end{abstract}
\keywords{galaxies: individual (Large Magellanic Cloud) ---
          galaxies: ISM --- 
          ISM: neutral hydrogen ---
          Magellanic Clouds --- 
          radio lines: galaxies}

\section{Introduction}

High resolution \hi\ observations of nearby galaxies allow the study
of many aspects of their dynamics, morphology and interstellar
physics. Measurement of the vertical \hi\ distribution and velocity
dispersion, for example, allows the mass-to-light ratio and dark
matter content of galaxy disks to be studied (Olling 1996). The shape
of the inner velocity field, using \hi\ as a tracer, can discriminate
between different halo models (de Blok et al. 2001). Also, the
relationship between \ion{H}{2} regions and \hi\ holes says much about
the evolution of galaxies and the propagation of star-forming regions
(Walter \& Brinks 2001). 

The LMC, as the nearest gas-rich galaxy to our own, has been the subject 
of several \hi\ studies (Luks \& Rohlfs 1992; McGee \& Milton 1966). The 
advantage of studying the LMC in \hi\ is that it is the nearest galaxy to
our own with a distance of 50$-$55 kpc (Feast 1991), it is presented nearly
face-on, and it is a very gas-rich and active star forming galaxy, so that
it allows for a detailed study of the structure, dynamics and the 
interstellar medium of a star-forming galaxy at close range. The early 
single-dish observations revealed the existence of expanding \hi\ shells 
in the LMC (Westerlund \& Mathewson 1996; McGee \& Milton 1966). A very
spectacular example of such a shell associated with the major star-forming 
region Constellation III was described by Dopita, Mathewson, \& Ford (1985).

However, its location in the far south means that high resolution studies 
had to await the advent of the Australia Telescope Compact Array\footnote
{The Australia Telescope Compact Array is funded by the is funded by the 
Commonwealth of Australia for operation as a National Facility managed by 
CSIRO.}(ATCA). The advent of ATCA finally allowed us to take full advantage 
of the aperture synthesis technique in southern hemisphere observations, and 
our recent high-resolution \hi\ survey of the Large Magellanic Cloud (LMC) 
revealed that the structure of the neutral atomic interstellar gas to be 
dominated by numerous shells and holes as well as complex filamentary 
structure (Kim et al. 1998). Its huge angular scale ($\sim$8$^{\circ}$ for
the inner disk), required effective observing techniques and mosaicing 
methods (Staveley-Smith et al. 1997), and required the development of new 
deconvolution algorithms (Sault, Staveley-Smith, \& Brouw 1996).

At the largest scales, understanding the interaction of the LMC 
with the Galaxy and the Small Magellanic Cloud (SMC) is important (Putman 
et al. 1998; Weinberg 2000). Also important is knowing the viewing geometry 
of the disk of the LMC (van der Marel \& Cioni 2001), its mass and dark 
matter content (Kim et al. 1998; Alves \& Nelson 2000) and the origin of 
the off-center bar (Zhao \& Evans 2000; Gardiner, Turfus, \& Putman 1998). 
At intermediate scales, the origin of the supergiant shells is tantalizing. 
Are supernovae and stellar winds (Dopita, Mathewson, \& Ford 1985; Chu \& 
Mac Low 1990) sufficient, or are other factors such as instabilities in the
ISM (Wada et al. 2000), high-velocity cloud collisions (Tenorio-Tagle \& 
Bodenheimer 1980), or gamma ray bursts (Efremov, Ehlerova, \& Palous 1999) 
important? At smaller scales, the structure of the ISM, the structure of 
photo-dissociation regions and the detailed feeding and feedback of 
star-forming regions are important for understanding their evolution.

In this paper, we perform the last step of our high-resolution \hi\ survey of 
the LMC by combining the already-made ATCA observations (Kim et al. 1998) with 
the relatively newer observations made with the multibeam receiver (Staveley-Smith 
et al. 2002) on the Parkes telescope. This step is 
necessary in order to provide a single-dish data cube containing structure over 
the full range of spatial frequencies on the sky. The ATCA images, on their 
own, may not provide reliable \hi\ masses for objects extended over more than
10\arcmin\ -- 20\arcmin\ as the minimum ATCA baseline for these observations
was 30m, for which the fringe spacing is $\lambda/D=24\arcmin$ (though
mosaicing extends the effective fringe spacing to $\sim 36\arcmin$ - see 
Holdaway 1999). Similarly, Parkes \hi\ images have a resolution of 14\arcmin
-- 16\arcmin, depending on the details of the gridding of the data, and are 
therefore not useful for examining structures smaller than $\sim25\arcmin$. 
Several examples of combination of aperture synthesis and single-dish
data can be found in the literature: Ye \& Turtle (1991) and Stanimirovi\'{c} et
al. (1999) utilize image-plane combination, whereas Bajaja \& van Albada (1979), 
and Wilner \& Welch (1994) utilize Fourier-plane combination. In this paper, we 
further discuss and develop data combination techniques, and use the most suitable 
to merge the \hi\ data for the LMC.

In \S~2 in this paper, we summarize the existing \hi\ data sets from the ATCA 
and Parkes telescopes and describe the method for linearly combining 
synthesis and single-dish observations. In \S~3, we present the
combined data set in the form of spatial-spatial map integrated over the
full velocity range, velocity channel maps, and spatial-velocity maps at
forty galactic longitude. In \S~4, we confirm the \hi\ giant and supergiant
shell candidates selected from the ATCA data using the new fully-sampled data,
and then compare the size distribution of \hi\ shells with the size distribution
of discrete \hi\ clouds in the LMC. In \S~5, we give a summary of this paper.

\section{Observation}
The Australia Telescope Compact Array (ATCA) has been used in mosaic mode 
to survey a region 10$^{\circ}$ $\times$ 12$^{\circ}$ covering the LMC and 
at an angular resolution of 1\farcm0, corresponding to a spatial resolution 
of 15 pc in the LMC. The detailed observations and data reductions are described 
in Kim et al. (1998).
The efficient observing techniques and mosaicing methods with the ATCA are 
described in Staveley-Smith et al. (1997).

The observing band was centered on 1.419 GHz, corresponding to a central 
heliocentric velocity coverage from $-$33 to $+$627 \kms\ , at a velocity 
resolution of 1.65 \kms\ . Imaging of the 
observed visibility data has been performed using the standard linear 
mosaicing scheme (Sault, Staveley-Smith, and Brouw 1996). After 
deconvolution the final cube was constructed by convolving the maximum 
entropy model with a circular Gaussian beam of FWHM 1$'$ $\times$ 1$'$ and 
adding the deconvolution residuals back in. The RMS noise of the final map, 
determined from the line-free parts of the final data cube, is 31 mJy 
beam$^{-1}$. The corresponding brightness temperature sensitivity is 2.5 K, 
which is equivalent to a column density sensitivity of 1.6 $\times$ 
10$^{19}$ cm$^{-2}$ per 1.649 km s$^{-1}$ channel.

\subsection{Parkes \hi\ Observations}

To provide the complete map at low spatial resolution to complement the 
ATCA data, observations were taken with the inner 7 beams of the Parkes 
Multibeam receiver (Staveley-Smith et al. 1996) on 1998 December 13 to 17 
(Staveley-Smith et al. 2001). The receiver was scanned across the LMC in 
orthogonal east-west and north-south great circles and the receiver was 
continuously rotated such that the rotation angle was always at $19\fdg1$ 
to the scan trajectory, thus ensuring uniform 
spatial sampling of the sky. The area covered was $13\arcdeg$ by $14\arcdeg$
in RA and Dec., respectively, and centered on $05^{\rm h} 20^{\rm m}$, 
$-68\arcdeg 44\arcmin$ (J2000). In a single scan, the spacing between 
adjacent tracks is 9\farcm5, which is smaller than the mean FWHP beam width 
of $14\farcm1$, but greater than the Nyquist interval ($\lambda/2D$) of 
$5\farcm7$. Therefore, six scans are interleaved in each of the principal 
scan directions, resulting in a final track spacing of $1\farcm6$. In total,
$12\times 6$ RA scans and $11\times6$ Dec. scans were made. Seven scans 
were dropped or edited out due to drive problems, leaving a total of 131 
scans consisting of a total of 29 hours of on-source integration on each of 
7 beams. The average integration time per beam area is 360 seconds for both 
polarizations. 
 
The scan rate of the telescope was $1\fdg0$ min$^{-1}$ and the correlator 
was read every 5 seconds. Therefore, the beam was slightly broadened in the 
scan direction to $14\farcm5$. After averaging orthogonal scans, the 
effective beam width reduces to $14\farcm3$. The central observing frequency 
was switched between 1417.5 and 1421.5 MHz, again every 5 seconds. 
This allowed the bandpass shape to be calibrated without spending any time 
off-source. A bandwidth of 8 MHz was used with 2048 spectral channels in each
of two orthogonal linear polarizations. \hi\ emission from the LMC appeared 
within the band, at both frequency settings. After bandpass calibration, the 
data from both settings were shifted to a common solar barycentric reference
frame. The velocity spacing of the multibeam data is 0.82 km s$^{-1}$, but 
the final cube was Hanning-smoothed to a resolution of 1.6 \kms\ . The useful
velocity range in the final cube (i.e. after excluding frequency side lobes of
the LMC and the Galaxy, and band-edge effects) is $-66$ to 430 \kms\ . 
 
Bandpass calibration, velocity shifting and preliminary spectral baseline 
fitting were all done using the {\sc aips$^{++}$} {\sc LiveData} task. 
Subsequently, the data were convolved onto a grid of $4\arcmin$ pixels using
a Gaussian kernel with a FWHP of $8\farcm0$. This broadens the effective, 
scan-broadened, beam width of the inner 7 beams from $14\farcm3$ to about 
$16\farcm4$. Residual spectral baselines were removed by fitting polynomials
in the image domain ({\sc miriad} task {\sc contsub}). The multibeam data 
were calibrated relative to a flux density for PKS~B1934$-$638 of 14.9 Jy at
the observing frequency. The brightness temperature conversion factor of 
0.80 K Jy$^{-1}$ was established by an observation of S9 ($T_B=85$ K, 
Williams 1973). On the same scale, we measured a brightness temperature for 
pointing 416 in the SMC ($00^{\rm h}47^{\rm m}52.6^{\rm s}$, $-73\arcdeg 
02\arcmin 19\farcs8$, J2000) of $T_B$ = 133 K, compared with the 137 K measured 
by Stanimirovi\'{c} et al. (1999). The 3\% difference is probably due to the 
different characteristics of the feeds used in the two observations, and residual 
uncertainties in absolute bandpass calibration. The rms noise in the line-free 
region of the cube is 27 mK, which is close to the theoretical value.   

\subsection{Combining ATCA and Parkes Observations}

Several techniques are available to combine the interferometric and 
single-dish data. The data can be combined during a joint maximum entropy 
deconvolution operation. Alternatively the single-dish data can be used 
as a 'default' image in a maximum entropy deconvolution of the interferometer
data. Another possibility is to feather together (a linear merging process)
the single-dish and interferometer images. Stanimirovic et al. (1999) have 
used an approach where the interferometer and single-dish image are added 
together before deconvolution, and then deconvolution performed with a 
modified point-spread function. Tests by Stanimirovic et al. (1999) found 
that the different techniques produced quite similar results. Given this, 
and as we already had a deconvolved interferometric cube, and as the 
computational requirements to re-perform a deconvolution of these data is 
large, we have used an image feathering approach. 

The technique we have used is a variant of the approach described by Schwarz
\& Walker (1991). As the Parkes and ATCA images give accurate representations
of the LMC at short spacings and mid to long spacings respectively, a 
composite image can be formed by filtering out the short spacing data from 
the ATCA image, and then adding the Parkes image. This process is most 
easily visualized in the Fourier domain as in Figure \ref{f:fig1} shows 
the expected amplitude as a function of spatial frequency of a point 
source for our observations. The Fourier transform of the Parkes images 
were added to the final images with no weighting (i.e. 'natural' weight in 
interferometric nomenclature). The deconvolved ATCA data was also 
Fourier-transformed, but by down-weighting the lower spatial frequencies such 
that the combined weight of the Parkes and ATCA data was the same as the response 
to a 1.$'$0 Gaussian. The {\sc miriad} task {\sc immerge} was used. 

Before combining, the Parkes image was interpolated onto the same 
coordinate grid as the ATCA mosaicked image. Also the residual primary 
beam attenuation remaining in the ATCA image was applied to the Parkes 
image (the mosaicing process we have used does not perform full primary 
beam correction when this would result in excessive noise amplification).
In order to perform the combination of the Parkes and ATCA observations, 
we need to ensure that the flux calibration between the two data types 
are consistent. Ideally we would like to find the ratio of the flux density
of an unresolved point source in the field. We have estimated this 
calibration factor by examining data in the Fourier plane between 21 and 
31 meters -- data in this annulus is well measured by the Parkes and 
mosaicked ATCA observations. After tapering the ATCA data to the same 
resolution as the Parkes data, we found a scale factor of 1.3 minimized 
the $L_1$ difference between the interferometer and single-dish Fourier 
components (the real and imaginary parts of the data where treated as 
distinct measurements in the fitting). Note that the scale factor (and 
indeed the entire feathering process) requires a good estimate of the 
resolution of the Parkes image. 
The effective beam size of 16.9 arcmin was adopted as it gave a scale 
factor independent of spatial frequencies (Stanimirovic et al. 1999). 

The resolution of the combined \hi\ image of the LMC is 1.$'$0 which 
is the same as for the ATCA interferometer map. The RMS noise of the 
combined map, determined from the line-free parts of the final data 
cube, is $\sim$19 mJy beam$^{-1}$. This corresponds to a brightness 
temperature sensitivity of $\sim$ 2.4 K.

\section{DATA PRESENTATION}

To assist the comparison, we display in Figure 2 the individual channel maps
from the combined \hi\ ATCA map with the Parkes single dish map. This figure 
can be compared with Figure 2 of Kim et al. (1998). The individual channel 
maps have a velocity resolution of 1.649 \kms\ and cover a velocity range 
of $V_{HEL}$=205 \kms\ to $V_{HEL}$=334 \kms\ . However, the \hi\ emission 
is detected mostly in the velocity range of $V_{HEL}$=190--387 \kms\ . The
peak \hi\ surface brightness image and the column density image of the LMC 
are shown in Figure 3 and Figure 4. The peak brightness temperature is 136.7 K
at RA=05$^{\rm h}$40$^{\rm m}$43$^{\rm s}$, DEC = $-$69$^{\circ}$48$'$49.7$''$
(J2000). The peak column density of 7.3$\pm$0.3 $\times$10$^{21}$ cm$^{-2}$ in 
the $V_{HEL}$=225$-$310 \kms\ assuming that the \hi\ is optically thin at this 
point. 

There is a remarkable correspondence between the features of this map and the 
\hi\ emission obtained from the previous ATCA survey (Kim et al. 1998). For 
example, the spiral structure is clearly seen in the individual channel maps 
(Figure 2b) and the peak \hi\ surface brightness map (Figure 3) as well 
as its integrated map (Figure 4). In contrast to the optical image of the 
LMC (Kim et al. 1999, Figure 5a), both ATCA map and ATCA$+$Parkes combined map
show that the \hi\ distribution is uniform and no bar feature corresponding to 
the optical one. The huge \hi\ hole of diameter about 1.2 kpc resides between 
the southern (de Vaucouleurs \& Freeman 1972) and northern spiral arm. 

The large-scale distribution of hydrogen in 30 Dor region and the south of 
30 Dor region in the coordinate range 05$^{h}$ 49$^{m}$ $<$ RA $<$ 05$^{\rm 
h}$ 36$^{\rm m}$, $-$73$^{\circ}$ 00$'$ $<$ DEC $<$ $-$68$^{\circ}$ 30$'$ 
displays two large sheets of gas having a relative difference in the 
line-of-sight velocity about 40 \kms\ but with relatively small internal 
velocity dispersions.
The general shape of this feature seen in the 
Position-Velocity ($P-V$) diagram (Figure 5) can be explained as the disk 
(D-) component and a second surface (L-component), discussed by Luks and 
Rohlfs (1992), which is possibly a region affected by the ram-pressure 
associated with the motion of the LMC through the outer halo of our Galaxy 
(Kim et al. 1998). However, many arclike structures seen in the $P-V$ diagram 
are likely to originate from expanding shells (Kim et al. 1999), although 
it is very difficult to distinguish between the components from expanding 
shells and intrinsic separate disk components (e.g., the L- and D- components) 
as a result of combined action of multiple expanding shells and local random 
motion. 

\section{SIZE SPECTRUM OF \hi\ SHELLS AND CLOUDS}

The structure of the neutral atomic ISM in the LMC shows a complex distribution
of \hi\ emission, which is chaotic with hundreds of clouds, shells, arcs, rings,
and filaments. The shell-like structures seen in the ATCA map still dominate the
structure of the neutral atomic ISM in the LMC revealed by the ATCA$+$Parkes combined
map. The previous visual survey of \hi\ shell candidates chosen from the ATCA \hi\ 
data cube (Kim et al. 1999) has been re-investigated from the current combined 
data set. Here we confirm that the \hi\ supergiant shells reported in Kim et al. 
(1999) are indeed seen in the combined data sets and presented in Figure 6 in this
paper. Figure 6 is the same as Figure 2 of Kim et al. (1999). 

The physical parameters of the individual supergiant shells are summarized in 
Table 1. The mean column density of neutral hydrogen in the LMC is in the order 
of 2.8 $\times$ 10$^{21}$ cm$^{-2}$. Assuming a mean column density is 
distributed uniformly over the disk thickness of $\Delta$ = 360 pc (Kim et 
al. 1999), we estimate a mean gas particle density of $n_H$ $\sim$ 2 
cm$^{-3}$. We may estimate the amount of kinetic energy of the \hi\ gas 
associated with the expanding \hi\ shells, using their derived sizes and 
expansion velocities. A total predicted kinetic energy of the interstellar 
gas associated with \hi\ expanding supershells is 3.9 $\pm$ 1.3 $\times$ 
10$^{53}$ ergs over the mean dynamical age of the expanding shell $\sim$ 6 
Myr. This result is remarkably similar to the total kinetic energy deposited
from the stellar winds $\sim$ 4.3 $\times$ 10$^{53}$ ergs over 6 Myr, 
referring the results derived from a total ionizing flux of $\sim$ 6.7 $\times$ 
10$^{51}$ photons s$^{-1}$ for UV sources in the LMC (Smith et al. 1987) and 
using the relationship between the ionizing photon flux and the stellar wind 
mechanical luminosity, $L_w/N_c = 3.2 \times 10^{-13}$ ergs photon$^{-1}$ 
(Wilson 1983).

\begin{table*}[h]  
\tablenum{1}
\center
\caption{List of positions, radii and heliocentric velocities for the \hi\ 
supergiant shells (SGSs) identified in the LMC.}
\vspace*{0.1cm}
\begin{tabular}{l@{\hspace{4mm}}cccc@{\hspace{4mm}}c@{\hspace{4mm}}c@{\hspace{
4mm}}
c@{\hspace{4mm}}c@{\hspace{4mm}}c}
\hline\hline
Shell  & RA & DEC & Shell  & Expanding & Heliocentric \\
       &    &     & Radius & Velocity  & Velocity \\
  & \multicolumn{2}{c}{(J2000)} & ($'$) & (\kms) & (\kms) & \\
\hline
SGS1 & 04:58:36 & $-$ 73:33:57 & 24.3$\times$22.7 & 19.0 & 253 \\
SGS2 & 04:58:30 & $-$ 68:39:29 & 18.7$\times$18.7 & 17.0 & 272 \\ 
SGS3 & 04:59:41 & $-$ 65:44:43 & 26.9$\times$26.9 & 15.0 & 294 \\
SGS4 & 05:02:51 & $-$ 70:33:15 & 36.7$\times$28.3& 23.0 & 241 \\
SGS5 & 05:04:08 & $-$ 68:31:55 & 36.6$\times$36.6 & 13.0 & 292 \\
SGS6 & 05:13:58 & $-$ 65:23:27 & 49.0$\times$37.5 & $-$ & 300 \\
SGS7 & 05:22:42 & $-$ 66:05:38 & 32.7$\times$11.2 & $-$ & 305 \\
SGS8 & 05:23:05 & $-$ 68:42:49 & 15.0$\times$15.0 & 22.5 & 271 \\
SGS9 & 05:25:46 & $-$ 71:09:16 & 30.7$\times$30.7 & $-$ & 235 \\
SGS10 & 05:30:45 & $-$ 68:04:30 & 32.7$\times$26.5 & $-$ & 284 & \\
SGS11 & 05:31:33 & $-$ 66:40:28 & 38.7$\times$37.7 & 36.0 & 306 \\
SGS12 & 05:30:26 & $-$ 69:07:56 & 41.1$\times$29.0 & $-$& 269 \\
SGS13 & 05:30:30 & $-$ 68:48:28 & 14.5$\times$21.2 & $-$ &    \\
SGS14 & 05:34:33 & $-$ 66:20:15 & 15.4$\times$15.4 & $-$ & 305 \\
SGS15 & 05:34:44 & $-$ 68:44:36 & 17.9$\times$17.9 & 24.0 & 279 \\
SGS16 & 05:36:54 & $-$ 68:27:46 & 15.3$\times$15.3 & 18.0 & 269 \\
SGS17 & 05:40:26 & $-$ 68:18:58 & 33.3$\times$26.7 & $-$ & 292 & \\
SGS18 & 05:41:00 & $-$ 71:15:18 & 30.0$\times$26.7 & $-$ & 223 \\
SGS19 & 05:41:27 & $-$ 69:22:23 & 26.0$\times$26.0 & $-$ & 259  \\
SGS20 & 05:46:49 & $-$ 70:02:32 & 24.7$\times$24.7 & $-$ & 259  \\
SGS21 & 05:44:53 & $-$ 66:28:29 & 13.0$\times$13.0 & 19.0 & 299 \\
SGS22 & 05:46:16 & $-$ 68:22:02 & 14.7$\times$14.7 & 21.5 & 303 \\
SGS23 & 05:51:16 & $-$ 67:37:18 & 41.3$\times$41.3 & 23.0 & 296 & \\
\hline
\end{tabular}
\end{table*}

The distribution by number of the \hi\ shells as a function of their radius
is presented in Figure 17 of Kim et al. (1999). In the range 100 -- 1000 pc, 
the data are consistent with a power-law distribution of slope $s=-1.5\pm 0.4$.
In order to compare the size spectrum of the \hi\ shells in the LMC with a size 
spectrum of \hi\ clouds, we investigate the \hi\ cloud candidates from the 
ATCA$+$Parkes combined data cube. It is well known that the interstellar 
medium, atomic as well as molecular, is distributed in a hierarchical ensemble 
of clouds. Such {\it clouds} or {\it clumps} could be a condensation formed 
during the thermally unstable cooling. The gas may initially be in either an 
atomic or a molecular state depending on the local physical conditions (density, 
excitation temperature, etc.). As the volume densities increase, the gas phase 
turns to molecular. Recent CO($J$=1$-$0) and CO($J$=2$-$1) study of the two 
nearly face-on galaxies NGC 628 and NGC 3938 shows that the velocity dispersion 
is remarkably constant with radius, 6 \kms\ for NGC 628 and 8.5 \kms\ for NGC 3938, 
and of the same order as the \hi\ velocity dispersion (Combes \& Becquaert 1997). 
The similarity of the CO and \hi\ dispersions suggests that the two components are 
well mixed, and are only two different phases of the same kinematical gas component.
The position of \hi\ clumps can be well matched with CO emission maps (Cohen et al. 
1988; Israel et al. 1993; Fukui et al. 1999). A large fraction of the \hi\ clumps 
have a detectable 100 $\mu$m emission. Similarly, where the \hi\ clumps 
are more intense, and the clumps are associated with CO clouds. The large-scale 
association between \hi\ and CO is clearly of central interests for studies of cloud 
and star formation in galaxies (Elmegreen \& Elmegreen 1987). 

Shapes of \hi\ clouds are difficult to define and the identification of clouds is 
still a subjective issue. However, majority of \hi\ clouds can be redefined as 
\hi\ {\it clumps} at small scales in either larger sheets or filaments abound 
rather than spherical blobs. 
The present \hi\ aperture synthesis survey of the LMC is particularly well 
suited to reach the statistical characteristics of \hi\ {\it clouds} or 
{\it clumps}. Since the distance of the LMC is known as 55 kpc (Feast 1991)
and the LMC is nearly face-on disk galaxy so that confusion along the line 
of sight is negligible. We have identified and catalogued \hi\ clouds in 
the LMC by defining a cloud to be an object composed of all pixels in right
ascension, declination, and velocity that are simply connected and that 
lie above the threshold brightness temperature (Scoville et al 1987; Lee et al. 
1990; Lee et al. 1997). We applied this method rather than gaussian clumping 
method as the other may generate all gaussian clumps which are not realistic in 
general. In fact, most of clouds do not have gaussian profiles, and especially 
it is true for HI clouds. Ideally, one would like to define clouds with a 0 K
threshold temperature.  However, low threshold temperatures are impractical
in view of the noise level in the spectra and more importantly because of
the blending of adjacent clouds which often occurs in crowded regions. We have 
found \hi\ clouds or clumps using the automatic clump identification code with 
three thresholds of the brightness temperature, $T_B$ =16 K ($\approx$5$\times$$T_{RMS}$), 
32 K, 64 K. Final selection of the \hi\ clouds has been made with high-temperature 
thresholds in order to reduce the blending of emission from unrelated clouds.

The distribution by number of the \hi\ clouds as a function of their size 
is presented in Figure 7. The derived sizes of clouds are distributed in a 
wide range of scales 20 $-$ 400 pc. The sizes are computed as the square root of 
the area. Peak of their size distribution of \hi\ clouds or clumps resides in 20 $-$ 
30 pc. However, the effective synthesis beam size 1$'$ limits the size distribution of 
the smallest \hi\ clouds. The data are consistent with a power-law distribution with 
slope of between $s$=$-$0.9 $\pm$0.2 and $s$=$-$1.1 $\pm$0.2. 
Compared to the slope found from the size spectrum of CO clouds in the inner 
Galaxy, $-$2.43$\pm$0.05 (Solomon et al. 1987); Ophiuchus, $-$2.20$\pm$0.06; 
Rosette, $-$2.26$\pm$0.08; Maddalena-Thaddeus, $-$2.46$\pm$0.20; M17, $-$2.09$
\pm$0.18 (Elmegreen \& Falgarone 1996), the power-law distribution of the size 
spectrum of \hi\ clouds is flatter than the typical size spectrum of CO clouds 
in the Milky Way. 

The measured perimeter $P$ and measured enclosed area $A$ of each \hi\ cloud 
in a log-log plot (Figure 8) gives a set of points lying along with a slope of 
$D/2$ = 0.73 $\pm$ 0.1. The relation between area and perimeter of each 
identified \hi\ cloud, $P$ $\propto$ $A^{D/2}$, can determine the fractal 
dimension $D$=1.47 $\pm$ 0.2 of a cloud boundary (Vogelaar \& Wakker 1994; 
Williams, Blitz, \& McKee 2000). The measured fractal dimension of \hi\ clouds
in the LMC is a similar dimension, $D$ $\pm$1.4, found in many studies of the
molecular ISM (Falgarone et al. 1991; Williams, Blitz, \& McKee 2000).  
For clouds identified with different thresholds of the brightness temperature,
the fractal dimension $D$ found from the relation between area and perimeter 
is invariant as shown in Figure 8. 

We find that a previous analysis of the holes and shells in the \hi\ distribution 
shows the same power-law behavior with that of clumps. The implication of the result 
is that the formation of the filament-like or shell-like structures, as well as 
clumps could be formed with the same power-law for dependence on the mass of gas in 
the galaxy and the stellar energy feedback. Satisfactory theoretical explanation needs 
to be developed for this observational fact.

\section{Summary}

We present the merged images of the LMC from the \hi\ Parkes multibeam 
observations with the aperture synthesis mosaic made by combining data from 
1344 separate pointing centers using the Australia Telescope Compact Array 
(ATCA). The images are constructed by a Fourier-plane technique and sensitive 
to structure on scales of 15 pc (for an LMC distance of 55 kpc) upwards. The 
addition of total power data reveals the plateau of diffuse \hi\ emission. We 
find that total power data is important in recovering the true source 
structures even linear mosaicing technique recovers the 'missing' short 
spacings in the $uv$ plane (Ekers \& Rots 1979; Cornwell 1988). 
The structure of the neutral atomic ISM in the LMC reveals the clumpiness of 
the \hi\ distribution over the whole of the LMC. We have identified and 
catalogued \hi\ clouds in the LMC by defining a cloud to be an object composed 
of all pixels in right ascension, declination, and velocity that are simply 
connected and that lie above the threshold brightness temperature. 
The power-law distribution of the size spectrum of \hi\ clouds is similar to 
the typical size spectrum of the holes and shells in the \hi\ distribution. 

\begin{acknowledgements}
We are indebted to the other team members of ATCA HI mosaic project Mike Kesteven, 
and Dave McConnell. We thank Bruce Elmegreen and Martin White for interesting 
discussions. We appreciate anonymous referee for improvement of this paper.

\end{acknowledgements}

\begin{figure}
\figurenum{1}
\caption{A representation of the feathering process in the Fourier domain.
This gives the expected visibility function of a point source from the
Parkes and ATCA observations. The feathering process filters out the short
spatial frequencies of the ATCA data, so that the sum of this and the
Parkes data give a good representation of the object at all spatial 
frequencies.}
\label{f:fig1}
\end{figure}

\begin{figure}
\figurenum{2}
\caption{The individual channel maps for the \hi\ datacube in the LMC. The 
heliocentric velocity is marked at the top left in each panel. Each panel 
is the average of five adjacent channels of width 1.649 \kms\ giving a panel
spacing of 8.2 \kms\ . The pots cover most of the \hi\ emission in the LMC 
from 210$-$334 \kms\ . Black represents regions with the highest brightness
temperatures of 136.7 K; white represents 0 K.}
\end{figure}

\begin{figure}
\figurenum{2}
\caption{{\it continued.}}
\end{figure}

\begin{figure}
\figurenum{2}
\caption{{\it continued.}}
\end{figure}

\begin{figure}
\figurenum{2}
\caption{{\it continued.}}
\end{figure}

\begin{figure}
\figurenum{3}
\caption{The peak \hi\ surface brightness map for the LMC. The grey scale intensity 
range is 0 to 136.7 K. This map is sensitive to the small \hi\ clouds with the 
highest opacity along the line of sight. In such regions, the brightness temperature
will approach the spin temperature of the \hi\ . The image is similar to the ATCA-only
image of Kim et al. (1998) and emphasizes the filamentary, bubbly and flocculent 
structure of the ISM in the LMC.}
\end{figure} 

\begin{figure}
\figurenum{4}
\caption{The \hi\ column density image of the LMC from the combined ATCA and Parkes
data. The intensity range is 0 to 7.3$\pm$0.3$\times$ 10$^{21}$ H-atom cm$^{-2}$ in 
\hi\ column density.} 
\label{fig3}
\end{figure}

\begin{figure}
\figurenum{5}
\epsscale{0.9}
\caption{Declination Velocity images of the LMC. Each panel is a slice through the 
LMC data cube at pa=0$^{\circ}$ at the RA specified at the top of each panel. The 
RA separation of the slices is $\sim$2.0$^{\rm m}$ or $\sim$ 10 beamwidths, and the
width of each slice is 23$^{\rm s}$, or $\sim2$ beamwidths. Only a fraction of the 
data are shown. The grey scale intensity range is 0 to 136.7 K.}
\label{fig4}
\end{figure}

\begin{figure}
\figurenum{5}
\epsscale{0.9}
\caption{\it continued.}
\label{fig4}
\end{figure}

\begin{figure}
\figurenum{5}
\epsscale{0.9}
\caption{\it continued.}
\label{fig4}
\end{figure}

\begin{figure}
\figurenum{5}
\epsscale{0.9}
\caption{\it continued.}
\label{fig5}
\end{figure}

\begin{figure}
\figurenum{6}
\caption{Supergiant shells are overlayed on the peak brightness HI map.}
\label{fig5}
\end{figure}


\begin{figure}
\figurenum{7}
\caption{Histograms of \hi\ clump size for three thresholds of the brightness
temperature, $T_B$=16K, 32K, 64K.}
\label{fig7}
\end{figure}

\begin{figure}
\figurenum{8}
\caption{A log-log plot of the measured perimeter versus the measured area 
in units of pixels of \hi\ clump for three thresholds of the brightness 
temperature, $T_B$=16K, 32K, 64K.}
\label{fig8}
\end{figure}

\end{document}